\documentclass{appolb}
\usepackage{epsfig}
\usepackage{feynarts}
\usepackage{axodraw}

\newcommand{\msbar}{$\overline{\rm{MS}}$}

\newcommand{\lsim}
{\;\raisebox{-.3em}{$\stackrel{\displaystyle <}{\sim}$}\;}

\newcommand{\BC}{\begin{center}}
\newcommand{\EC}{\end{center}}
\newcommand{\BE}{\begin{equation}}
\newcommand{\EE}{\end{equation}}
\newcommand{\BEA}{\begin{eqnarray}}
\newcommand{\BEAnn}{\begin{eqnarray*}}
\newcommand{\EEA}{\end{eqnarray}}
\newcommand{\EEAnn}{\end{eqnarray*}}

\begin{document}
\title{Top Quark Physics at Colliders
\thanks{Presented at the final meeting of the European Network "Physics at
Colliders", Montpellier, September 25-28,2004.
This work has been supported by the European Community's Human
Potential Programme under contract HPRN-CT-2000-00149.
}%
}
\author{E.W.N. Glover
\address{Institute for Particle Physics Phenomenology, University of Durham,\\
Durham DH1~3LE, UK}
\and
F.~del \'Aguila, J.A.~Aguilar--Saavedra,  M.~Beccaria, S.~B\'{e}jar, 
A.~Brandenburg, J.~Fleischer,
J.~Guasch, T.~Hahn, W.~Hollik,
S.~Heinemeyer, S.~Kraml, A.~Leike, A.~Lorca, W.~Porod, S.~Prelovsek,
F.M.~Renard, T.~Riemann,
C.~Schappacher, Z.G.~Si,
J.~Sol\`{a}, P.~Uwer,  C.~Verzegnassi, G.~Weiglein and A.~Werthenbach
\address{Universitat~Autonoma~de~Barcelona,
Universitat~de~Barcelona,
CERN,
Univ.~of~Durham,
Univ.d~de~Granada,
DESY,~Hamburg,
INFN~Lecce,
Univ.~di~Lecce,
Instituto~Superior~Tecnico~Lisbon,
Univ.~de~Montpellier II,
Ludwig-Maximilians-Univ.~M\"unchen,
Max~Planck~Inst.~f\" ur Phys.~M\"unchen,
Shandong~Univ.,
INFN~Trieste,
Univ.~di~Trieste,
Paul~Scherrer~Institut~Villigen,
DESY,~Zeuthen,
Univ.~Z\"urich}
}
\maketitle
\begin{abstract}
We review some recent developments in top quark production and decay
at current and future colliders.
\end{abstract}
\PACS{12.15.Lk, 12.38.Bx, 13.66.Bc,  14.65.Ha}
  
\section{Introduction}
The detailed analysis of the dynamics of top quark production and decay 
is a major objective of experiments at the Tevatron, the LHC, and a possible
international linear $e^+e^-$ collider (ILC). 
A special feature of the top quark that
makes such studies very attractive is its large decay width,
 $\Gamma_t\approx 1.48$~GeV, which  serves as a cut-off for non-perturbative
effects in top quark decays. As a consequence 
{\em precise} theoretical predictions of
cross sections and differential distributions involving top quarks 
are possible within the Standard Model and its extensions.
A confrontation  of such predictions 
with forthcoming high-precision data will lead
to accurate determinations of Standard Model parameters and possibly
hints of new phenomena. 

For more details on the general subject 
of top physics, we refer the reader to the recent collider studies 
\cite{Beneke:2000hk,
Aguilar-Saavedra:2001rg,
Abe:2001wn,Abe:2001gc} and 
 references therein.
In this talk, I review the joint contribution to top quark physics
made by the network.

\section{Top quark production at the
ILC~\cite{Fleischer:2003kk,Fleischer:2002kg,Fleischer:2002nn,Fleischer:2002rn,Hahn:2003ab}}

At the ILC, one of the most important {reactions will be
top-pair production well above the threshold} (i.e. in the
continuum region),
\begin{eqnarray}
  \label{eq:qqtt}
  e^+ ~+~ e^- \rightarrow t ~+~  \bar t~ \,\,.
\end{eqnarray}
Several hundred thousand events are expected, and the anticipated
accuracy of the corresponding theoretical
predictions should be around a few per mille. 
Of course, it is not only
the two-fermion production process (\ref{eq:qqtt}), with electroweak (EW) 
and
QCD radiative corrections to the final state that must be calculated with
high precision.
In addition, the decay
of the top 
quarks and a variety of quite different radiative corrections such as
real photonic bremsstrahlung and other non-factorizing contributions
to six-fermion production and beamstrahlung  have to be considered.  
New physics effects may also have to be taken into
account.

In \cite{Fujimoto:1988hu,Yuasa:1999rg},  the complete
$O(\alpha)$ corrections, including hard photon radiation, are calculated. The
virtual and soft photon corrections both in the Standard Model (SM)
and in the
minimal supersymmetric Standard Model (MSSM) are determined in 
\cite{Beenakker:1991ca,Hollik:1998md}, and (only) in the    SM in
\cite{Bardin:2000kn}.
At the time of the public presentation of the TESLA 
Technical Design Report~\cite{Aguilar-Saavedra:2001rg}.
detailed comparisons between these calculations had not been made.
For this reason, and to
 produce an event generator for the evaluation of experimental data,
the fortran code
{\tt topfit} has been written~\cite{Fleischer:2003kk,Fleischer:2002kg} 
which describes the fixed-order QED and electroweak one-loop corrections to
top pair production.

Top quark pair production from $e^+e^-$
annihilation at one-loop differs from light fermion production because
two new structures appear in the theoretical description that are a
consequence of the fact that the top mass is not negligible.  To
understand the origin of the extra structures, it is sufficient to
consider the theoretical expansion of a one-loop vertex coupling the top quark
pair to
either a photon or a $Z$.
In full generality, with CP-conserving interactions
one can identify the effective vertex~\cite{Beccaria:2000jz}
\begin{equation}
\Gamma^X_{\mu}=-e^X\left[\gamma_{\mu}(g^X_{Vt}-g^X_{At}\gamma^5)+{d^X\over
m_t}(p-p')_{\mu}\right]
\label{3forms}
\end{equation}
where $X=\gamma,Z$,  $e^{\gamma}=|e|$, $e^Z={|e|\over2s_Wc_W}$
and $p$, $p'$ represent the outgoing $t$, $\bar t$ momenta;
$g^X_{Vt},~g^X_{At},~d^X$ are $O(\alpha)$ one-loop contributions
which in general are
$q^2=(p+p')^2$ dependent. 
The new quantity $d^X$ enters because the top mass cannot 
be  neglected and appear
in the various theoretical expressions at one loop, making the overall
number of independent amplitudes of the process to increase from four
(in massless fermion production) to six. This is because the
\underline{three} independent coefficients of eq.~(\ref{3forms}) 
will be combined
with the \underline{two} independent coefficients ($g^X_{Vl},~g^X_{Al}$)
of the initial
(massless) lepton current. 

The one-loop corrections to $t\bar t$ production therefore consists of
evaluating these six form factors.
Typical one-loop vertex and box graphs contributing 
to the EW and QED corrections process are shown in
Fig.~1.
\begin{center}
\begin{figure}[tb!]
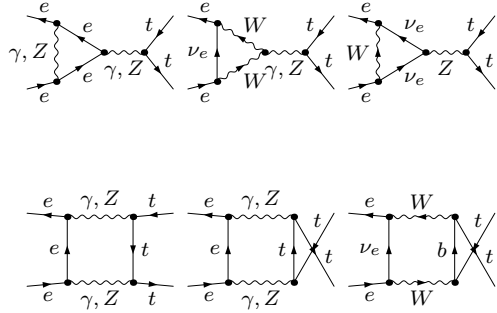

\label{fig:eett}
\unitlength=0.60bp%

\begin{scriptsize}
\begin{feynartspicture}(632,101)(4,1)
\FADiagram{}
\FAProp(0.,15.)(4.,14.)(0.,){/Straight}{-1}
\FALabel(2.37593,15.5237)[b]{$e$}
\FAProp(0.,5.)(4.,6.)(0.,){/Straight}{1}
\FALabel(2.37593,4.47628)[t]{$e$}
\FAProp(20.,15.)(16.,10.)(0.,){/Straight}{1}
\FALabel(17.2697,12.9883)[br]{$t$}
\FAProp(20.,5.)(16.,10.)(0.,){/Straight}{-1}
\FALabel(18.7303,7.98828)[bl]{$t$}
\FAProp(16.,10.)(10.5,10.)(0.,){/Sine}{0}
\FALabel(13.25,8.93)[t]{$\gamma, Z$}
\FAProp(4.,14.)(4.,6.)(0.,){/Sine}{0}
\FALabel(2.93,10.)[r]{$\gamma, Z$}
\FAProp(4.,14.)(10.5,10.)(0.,){/Straight}{-1}
\FALabel(7.58235,12.8401)[bl]{$e$}
\FAProp(4.,6.)(10.5,10.)(0.,){/Straight}{1}
\FALabel(7.58235,7.15993)[tl]{$e$}
\FAVert(4.,14.){0}
\FAVert(4.,6.){0}
\FAVert(16.,10.){0}
\FAVert(10.5,10.){0}

\FADiagram{}
\FAProp(0.,15.)(4.,14.)(0.,){/Straight}{-1}
\FALabel(2.37593,15.5237)[b]{$e$}
\FAProp(0.,5.)(4.,6.)(0.,){/Straight}{1}
\FALabel(2.37593,4.47628)[t]{$e$}
\FAProp(20.,15.)(16.,10.)(0.,){/Straight}{1}
\FALabel(17.2697,12.9883)[br]{$t$}
\FAProp(20.,5.)(16.,10.)(0.,){/Straight}{-1}
\FALabel(18.7303,7.98828)[bl]{$t$}
\FAProp(16.,10.)(10.5,10.)(0.,){/Sine}{0}
\FALabel(13.25,8.93)[t]{$\gamma, Z$}
\FAProp(4.,14.)(4.,6.)(0.,){/Straight}{-1}
\FALabel(2.93,10.)[r]{$\nu_e$}
\FAProp(4.,14.)(10.5,10.)(0.,){/Sine}{-1}
\FALabel(7.58235,12.8401)[bl]{$W$}
\FAProp(4.,6.)(10.5,10.)(0.,){/Sine}{1}
\FALabel(7.58235,7.15993)[tl]{$W$}
\FAVert(4.,14.){0}
\FAVert(4.,6.){0}
\FAVert(16.,10.){0}
\FAVert(10.5,10.){0}

\FADiagram{}
\FAProp(0.,15.)(4.,14.)(0.,){/Straight}{-1}
\FALabel(2.37593,15.5237)[b]{$e$}
\FAProp(0.,5.)(4.,6.)(0.,){/Straight}{1}
\FALabel(2.37593,4.47628)[t]{$e$}
\FAProp(20.,15.)(16.,10.)(0.,){/Straight}{1}
\FALabel(17.2697,12.9883)[br]{$t$}
\FAProp(20.,5.)(16.,10.)(0.,){/Straight}{-1}
\FALabel(18.7303,7.98828)[bl]{$t$}
\FAProp(16.,10.)(10.5,10.)(0.,){/Sine}{0}
\FALabel(13.25,8.93)[t]{$Z$}
\FAProp(4.,14.)(4.,6.)(0.,){/Sine}{-1}
\FALabel(2.93,10.)[r]{$W$}
\FAProp(4.,14.)(10.5,10.)(0.,){/Straight}{-1}
\FALabel(7.58235,12.8401)[bl]{$\nu_e$}
\FAProp(4.,6.)(10.5,10.)(0.,){/Straight}{1}
\FALabel(7.58235,7.15993)[tl]{$\nu_e$}
\FAVert(4.,14.){0}
\FAVert(4.,6.){0}
\FAVert(16.,10.){0}
\FAVert(10.5,10.){0}
\end{feynartspicture}
\end{scriptsize}

\unitlength=0.60bp%

\begin{scriptsize}
\begin{feynartspicture}(632,101)(4,1)
\FADiagram{}
\FAProp(0.,15.)(5.5,14.5)(0.,){/Straight}{-1}
\FALabel(2.89033,15.8136)[b]{$e$}
\FAProp(0.,5.)(5.5,5.5)(0.,){/Straight}{1}
\FALabel(2.89033,4.18637)[t]{$e$}
\FAProp(20.,15.)(14.5,14.5)(0.,){/Straight}{1}
\FALabel(17.1097,15.8136)[b]{$t$}
\FAProp(20.,5.)(14.5,5.5)(0.,){/Straight}{-1}
\FALabel(17.1097,4.18637)[t]{$t$}
\FAProp(5.5,14.5)(5.5,5.5)(0.,){/Straight}{-1}
\FALabel(4.43,10.)[r]{$e$}
\FAProp(5.5,14.5)(14.5,14.5)(0.,){/Sine}{0}
\FALabel(10.,15.57)[b]{$\gamma, Z$}
\FAProp(5.5,5.5)(14.5,5.5)(0.,){/Sine}{0}
\FALabel(10.,4.43)[t]{$\gamma, Z$}
\FAProp(14.5,14.5)(14.5,5.5)(0.,){/Straight}{1}
\FALabel(15.57,10.)[l]{$t$}
\FAVert(5.5,14.5){0}
\FAVert(5.5,5.5){0}
\FAVert(14.5,14.5){0}
\FAVert(14.5,5.5){0}

\FADiagram{}
\FAProp(0.,15.)(5.5,14.5)(0.,){/Straight}{-1}
\FALabel(2.89033,15.8136)[b]{$e$}
\FAProp(0.,5.)(5.5,5.5)(0.,){/Straight}{1}
\FALabel(2.89033,4.18637)[t]{$e$}
\FAProp(20.,15.)(14.5,5.5)(0.,){/Straight}{1}
\FALabel(18.2636,13.1453)[br]{$t$}
\FAProp(20.,5.)(14.5,14.5)(0.,){/Straight}{-1}
\FALabel(18.7364,8.14526)[bl]{$t$}
\FAProp(5.5,14.5)(5.5,5.5)(0.,){/Straight}{-1}
\FALabel(4.43,10.)[r]{$e$}
\FAProp(5.5,14.5)(14.5,14.5)(0.,){/Sine}{0}
\FALabel(10.,15.57)[b]{$\gamma, Z$}
\FAProp(5.5,5.5)(14.5,5.5)(0.,){/Sine}{0}
\FALabel(10.,4.43)[t]{$\gamma, Z$}
\FAProp(14.5,5.5)(14.5,14.5)(0.,){/Straight}{1}
\FALabel(13.43,10.)[r]{$t$}
\FAVert(5.5,14.5){0}
\FAVert(5.5,5.5){0}
\FAVert(14.5,5.5){0}
\FAVert(14.5,14.5){0}

\FADiagram{}
\FAProp(0.,15.)(5.5,14.5)(0.,){/Straight}{-1}
\FALabel(2.89033,15.8136)[b]{$e$}
\FAProp(0.,5.)(5.5,5.5)(0.,){/Straight}{1}
\FALabel(2.89033,4.18637)[t]{$e$}
\FAProp(20.,15.)(14.5,5.5)(0.,){/Straight}{1}
\FALabel(18.2636,13.1453)[br]{$t$}
\FAProp(20.,5.)(14.5,14.5)(0.,){/Straight}{-1}
\FALabel(18.7364,8.14526)[bl]{$t$}
\FAProp(5.5,14.5)(5.5,5.5)(0.,){/Straight}{-1}
\FALabel(4.43,10.)[r]{$\nu_e$}
\FAProp(5.5,14.5)(14.5,14.5)(0.,){/Sine}{-1}
\FALabel(10.,15.57)[b]{$W$}
\FAProp(5.5,5.5)(14.5,5.5)(0.,){/Sine}{1}
\FALabel(10.,4.43)[t]{$W$}
\FAProp(14.5,5.5)(14.5,14.5)(0.,){/Straight}{1}
\FALabel(13.43,10.)[r]{$b$}
\FAVert(5.5,14.5){0}
\FAVert(5.5,5.5){0}
\FAVert(14.5,5.5){0}
\FAVert(14.5,14.5){0}
\end{feynartspicture}
\end{scriptsize}
\caption{Typical graphs contributing to the weak and QED corrections to $e^+e^-
\to t\bar t$}
\end{figure}
\end{center}
The virtual corrections contain both ultraviolet (UV) and infrared (IR)
divergences and are treated by dimensional regularization. 
The  UV divergences  are eliminated by renormalization on the
amplitude level, while the  IR poles can only be eliminated at the
cross-section level by including the emission of soft photons. 

\begin{figure}[tb!]
\begin{center}
 \epsfig{file=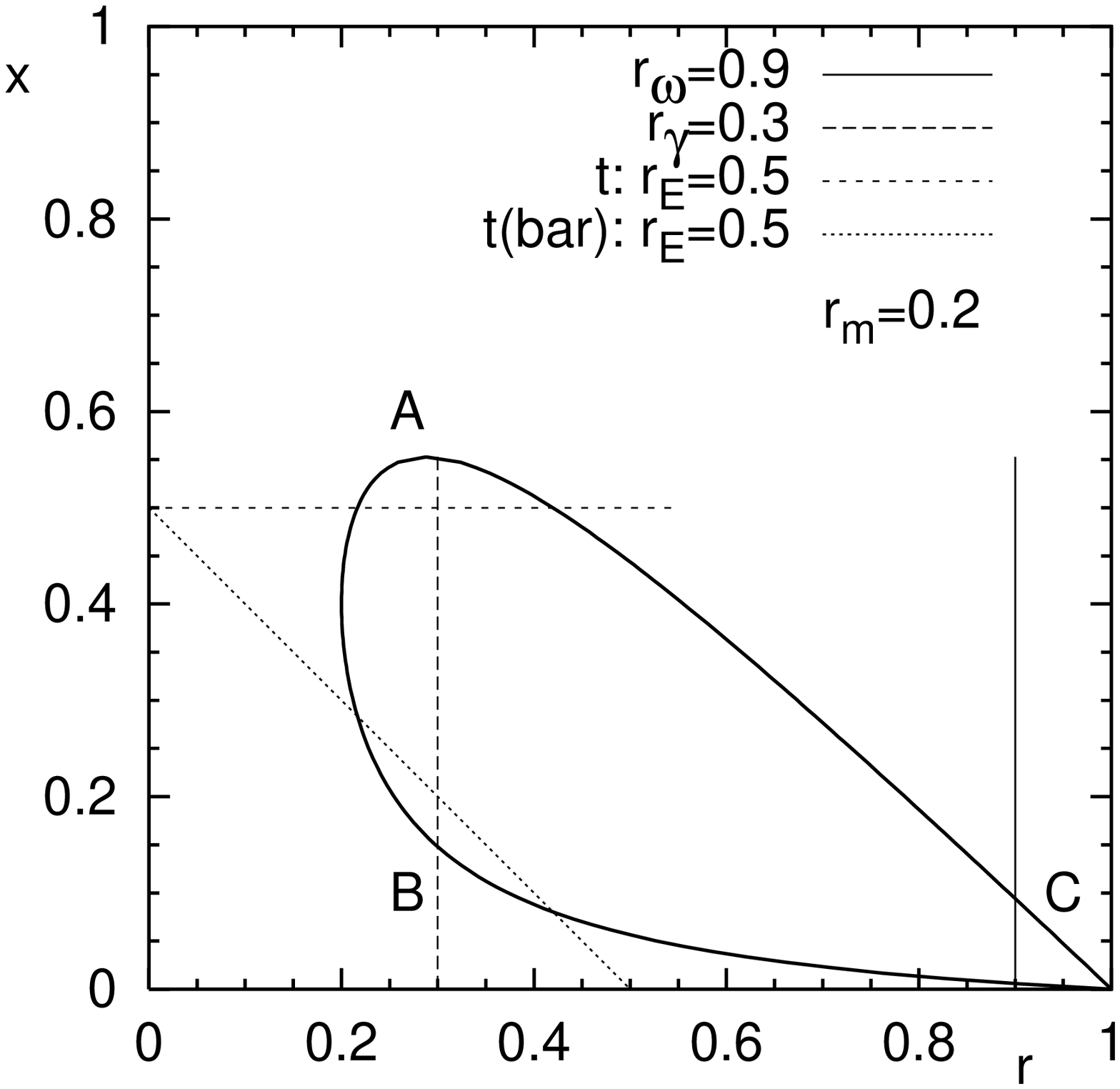,width=6.0cm}
 \epsfig{file=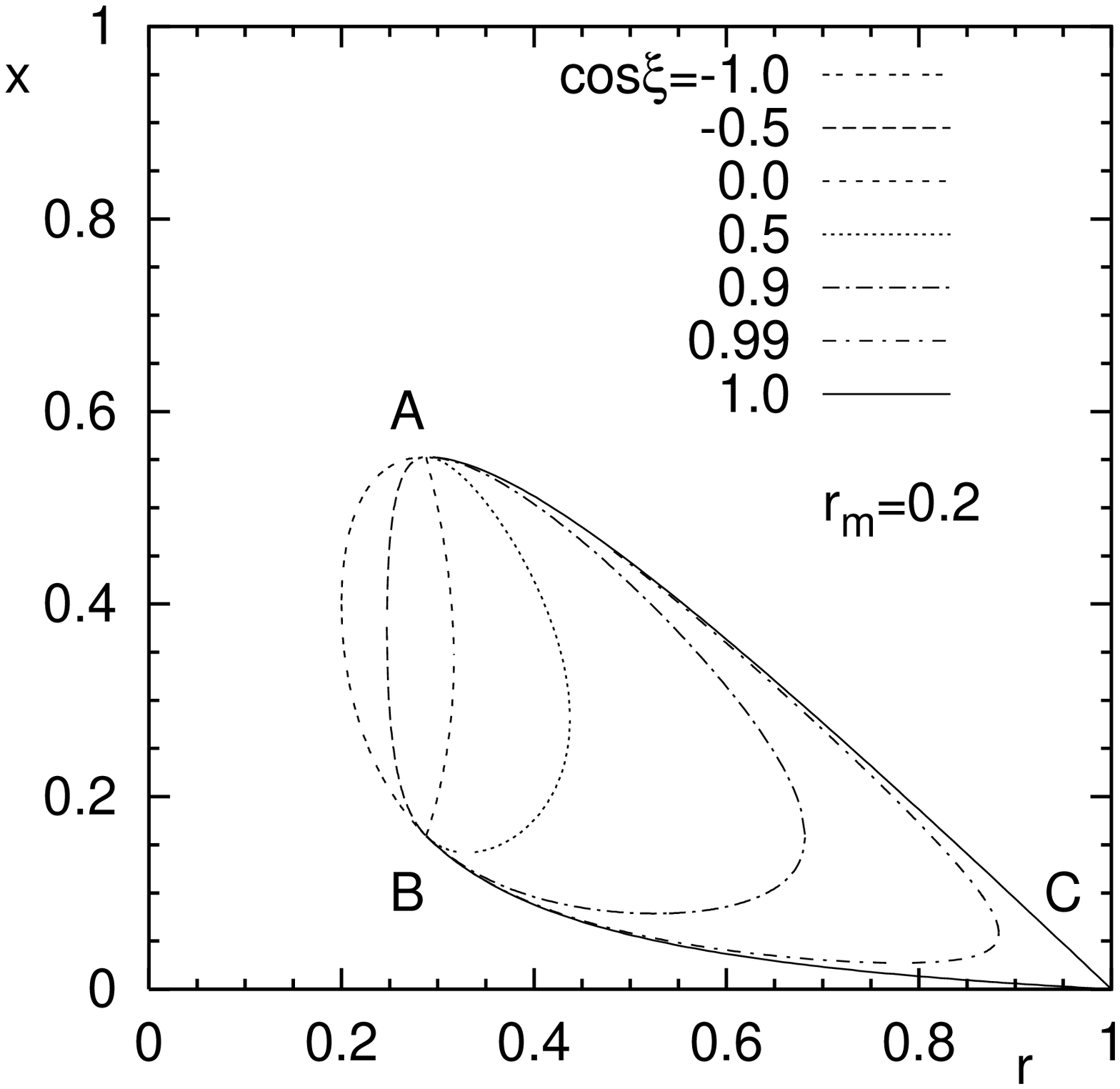,width=6.0cm}
\end{center}
\caption
{Phase space for 
$r_m=\frac{4 m_t^2}{s}=0.2$ (non-zero top mass).
Energetic cuts are also shown in (a)
$r_E=2E_{t}^{min}/\sqrt{s}$, 
$r_{\bar E}=2\bar E_{t}^{min}/\sqrt{s}$, 
$r_\omega=1-2\omega/\sqrt{s}=1-2E_{min}(\gamma)/\sqrt{s}$, 
$r_\gamma=1-2E_{max}(\gamma)\sqrt{s}$
while (b) shows
different values of the acollinearity
angle $\xi = \pi - \theta_{t\bar t}$.  Note that $\cos \xi = 1$ corresponds to
the elastic case.
\label{fig:phasecuts}
}
\end{figure}

The real radiation contribution is evaluated using a semi-analytical
integration approach with physically accessible observables as integration
variables.  This allows control over the numerical precision to  more than four
digits.   The phase space with three particles in the final state is
five-dimensional.  However, two of the angles are trivial and may be
integrated out so that, 
\begin{equation}
d\sigma = \frac{1}{(2\pi)^4}\frac{1}{2s\beta_0}|{\cal M}|^2~
\frac{\pi s}{16}\;\mbox{d}r\mbox{d}x\mbox{d}\cos\theta,
\label{space}
\end{equation}
where $\theta$ is the angle between the anti-top quark and the positron,
$x = 2 p_\gamma\cdot p_{\bar t}/s$ and $r = (p_t+p_{\bar t})^2/s$.
Cuts on the energies of the photon or top quarks, or cuts on the angles between
particles directly transfer into cuts on these variables.
The phase space is illustrated in Fig.~\ref{fig:phasecuts}.
%
The $t$ ($\bar t$) are at rest at points $A$ ($B$).
Soft photons are located at point $C$.   All phase space points away from $(r,x)
= (1,0)$ are finite and can be obtained numerically for any set of reasonable
cuts.  The soft photon contribution is analytically removed and combined with
the virtual graphs.

Numerical results from {\tt topfit} have been compared with two other groups.   First the 
virtual and soft photon contribution have been compared with results from
the Karlsruhe group~\cite{Fleischer:2002rn,Hahn:2003ab}.  The weak virtual corrections to
the angular distributions agree
to twelve digits, while the pure photonic corrections agree to at least eleven
digits.   Second, the hard photon corrections have been compared with results
from the GRACE group~\cite{Yuasa:1999rg}.  Depending on the observable,
agreement to three digits is generally obtained.

\begin{figure}[tb!]
\begin{center}
\hspace*{-1cm}
\mbox{\epsfysize=6.5cm\epsffile{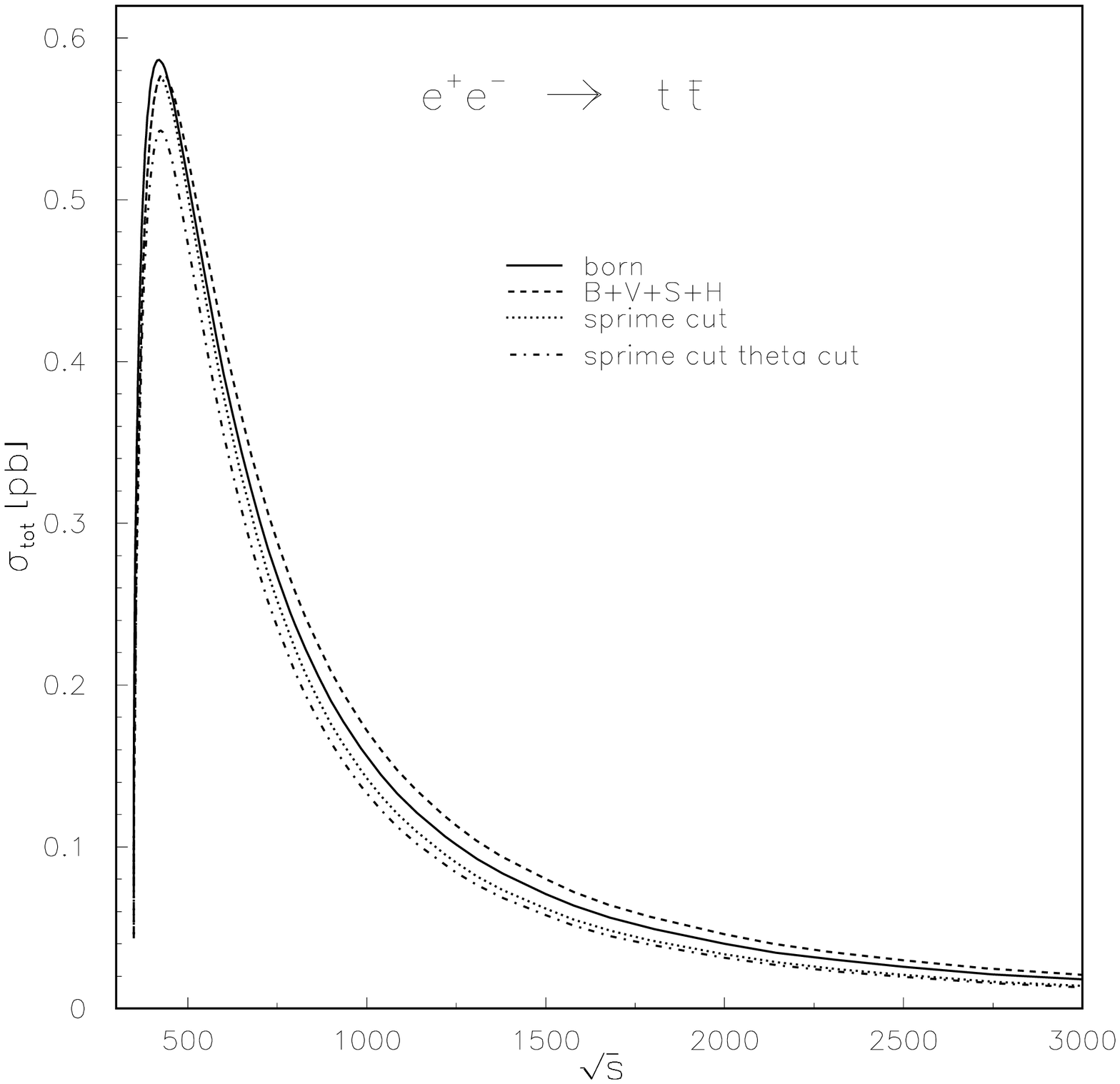}}
\mbox{\epsfysize=6.5cm\epsffile{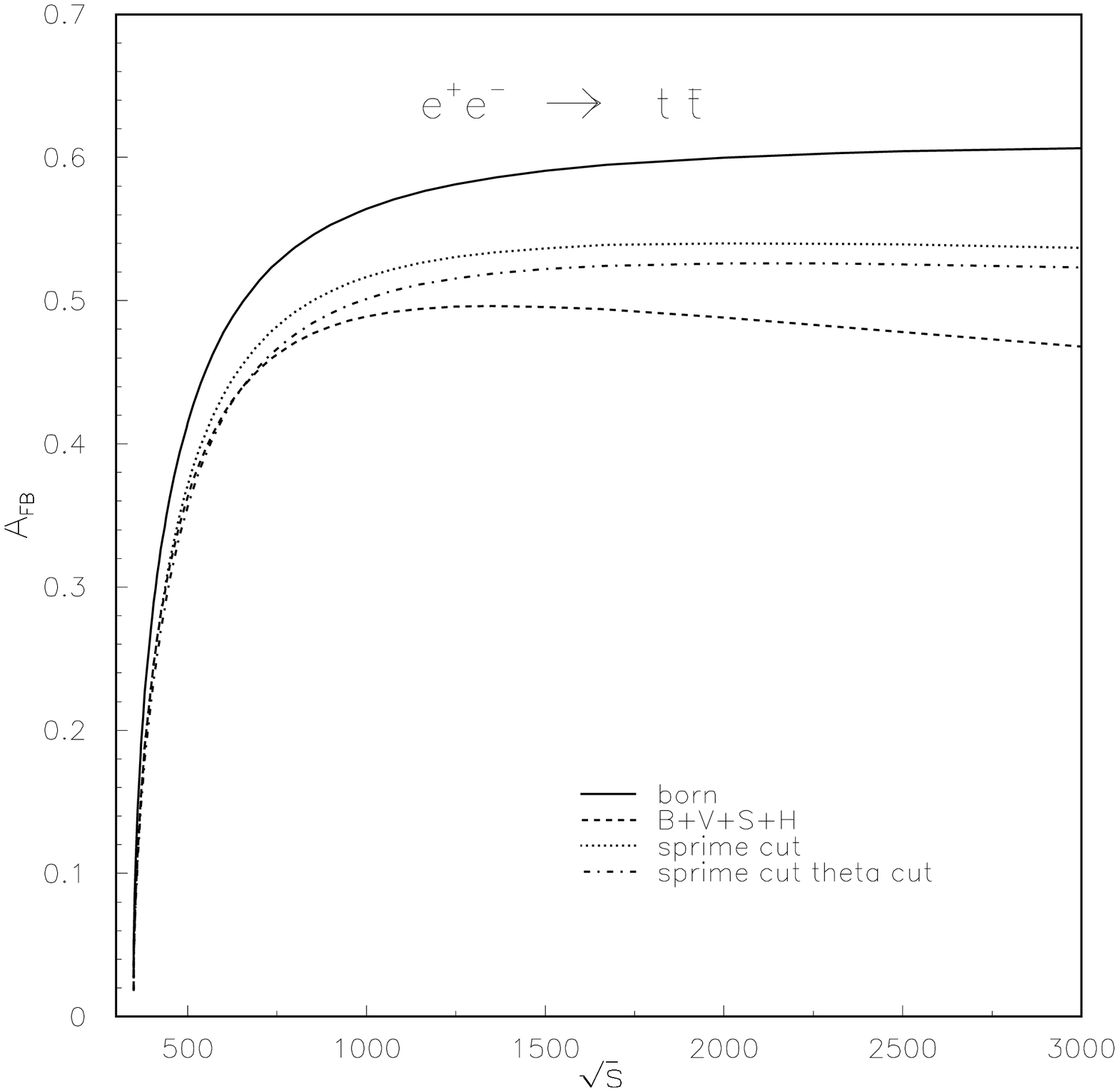}}
\vspace*{-1cm}
\end{center}
\caption[]{\label{fig:physres}
The (a) total cross-section and (b)
forward--backward asymmetry  for top-pair production as a function of $s$.
Born (solid lines), electroweak (dashed lines), electroweak with
$s'= 0.7 \, s$-cut (dotted lines)  and electroweak with  $s'=0.7 \, s$- and
$\cos\theta = 0.95$-cut (dash-dotted lines).}
\end{figure}
%

Fig.~\ref{fig:physres} shows the (a) total cross section and (b)
forward-backward asymmetry as a function of $\sqrt{s}$.\footnote{Note that this
is a fixed-order $\alpha$ calculation, i.e. no higher order corrections such as
photon exponentiation have been taken into account} The values of the input
parameters can be found in Ref.~\cite{Fleischer:2003kk}. The effects of
radiative corrections are more dramatic for top-pairs produced close to the
direction of the beam. For the ILC range of centre-of-mass energies, 
backward scattered top quarks give rise to slightly larger corrections to the
toral cross section
than
forward scattered ones~\cite{Fleischer:2002rn}. For higher energies this
effect is more or less washed out. This is not the case for the
forward--backward asymmetry.

In summary Ref.~\cite{Fleischer:2003kk} shows that at the ILC,
EW radiative corrections modify the differential (as well as the
integrated) top-quark observables
by more than the anticipated experimental precision of a few per mille.
The package {\tt topfit} provides the means to calculate those corrections and
allows predictions for various realistic cuts on the scattering angle as well
as on the energy of the photon.  The successful comparison with
Refs.~\cite{Fleischer:2002nn,Fleischer:2002rn} means that the technical
precision of {\tt topfit} is completely tested.

\section{Polarised top quark decay~\cite{Brandenburg:2002xr}}

In $e^+e^-$ collisions, top quarks are produced highly polarized, especially 
if one  tunes the polarization of the incoming beams, as possible e.g. at the
ILC collider~\cite{Aguilar-Saavedra:2001rg}.  At the LHC  the polarization of
top quarks  is  tiny due to parity and time reversal  invariance  of QCD.
However the spins of  $t$ and $\bar t$ are in general highly correlated.

The polarization  of the top quark  is transferred to the angular distribution
of its decay products through its weak, parity violating decays. 
If we consider a polarized  ensemble of top quarks at rest with polarization
vector ${\bf P},\ 0\le |{\bf P}|\le 1$,  the differential decay distribution
with respect to the angle $\vartheta$ between ${\bf P}$ and the direction
$\hat{\bf p}$ of a given decay product is given by,
\begin{eqnarray}\label{power}
\frac{1}{\Gamma}\frac{d\Gamma}{d\cos\vartheta}=\frac{1}{2}\left(1+
|{\bf P}|\kappa_p\cos\vartheta\right).
\end{eqnarray} 
In Eq.~(\ref{power}), ${\Gamma}$ is the partial width for the  corresponding
decay of unpolarized top quarks, and $\kappa_p$ is the so-called {\it spin
analysing power} of the final state  particle or jet under consideration. For
example, in the semileptonic decay $t\to l^+\nu_l b$, the charged  lepton
($b$-quark) has spin analysing power $\kappa_p=+1$ ($\sim -0.41$)  at the tree
level within the Standard Model.  In hadronic top decays $t\to b \bar{d} u$
(where $d (u)$ stands generically for $d,s\ (u,c)$),    the r\^ole of the
charged lepton is played by the  $\bar{d}$ quark.  However, the $\bar d$ quark
cannot be easily identified,  but with a 61\% probability is contained in the
least energetic light (i.e. non-$b$-quark)  jet. The spin analysing for the
least energetic jet is denoted by $\kappa_j$.

The QCD corrections to $\kappa_p$ for hadronic top decays 
are computed in Ref.~\cite{Brandenburg:2002xr}. 
These corrections are one 
ingredient in a full analysis of
top quark (pair) production and decay at next-to-leading
order in $\alpha_s$, both at lepton
and hadron colliders. They form part of the {\it factorizable}
corrections within the pole approximation for the top quark propagator. 
The QCD corrections for semileptonic
polarized top quark decays have been computed
in ref.~\cite{Czarnecki:1991}.

The size of the next-to-leading order (NLO) correction is defined as
\begin{eqnarray}\label{kappa_nlo}
\kappa_p \equiv \kappa_p^0[1+\delta_p^{QCD}]
+O(\alpha_s^2) ,
\end{eqnarray}
where $\kappa_p^{0}$ denotes the Born result.
Table~\ref{tab:nlo}
 shows that the top-spin analysing powers
of the final states in non-leptonic top quark decays receive
QCD corrections in the range $+1.4$\% to $-7.2$\%. 
The spin analysing power of jets is smaller than that of
the corresponding bare quarks.
This has to be contrasted
with the spin analysing power of the charged lepton in  
decays $t(\uparrow)\to
bl^+\nu_l$ where the QCD corrected result (for $m_b=0$) 
reads~\cite{Czarnecki:1991} $\kappa_l=1-0.015\alpha_s$, i.e. the correction
is at the per mille level.
 
\begin{table}[h]
\caption{QCD-corrected results for spin analysing powers.}
\begin{center}
\begin{tabular}{
|c|c|c|c|}\hline
& partons & jets, E-alg.  & jets, D-alg. \\  \hline
$\kappa_{\bar{d}}$ 
& $0.9664(7)$&  $0.9379(8)$ & 
$0.9327(8)$ \\ 
$\delta^{QCD}_{\bar{d}}$ [\%] & $-3.36\pm 0.07$ & $-6.21\pm 0.08$
 &  $-6.73 \pm 0.08$   \\ \hline  
$\kappa_{b}$ 
& $-0.3925(6)$& $-0.3907(6)$ & 
$-0.3910(6)$\\
$\delta^{QCD}_{b}$ [\%] & $-3.80\pm 0.15$ & $-4.24\pm  0.15$ & 
$-4.18  \pm 0.15$ \\ \hline  
$\kappa_{u}$ 
& $-0.3167(6)$&  $-0.3032(6)$ & 
$-0.3054(6)$ \\
$\delta^{QCD}_{u}$ [\%] & $+1.39\pm 0.19$ & $-2.93\pm 0.19$ & 
$-2.22\pm  0.19$   \\ \hline
$\kappa_j$ 
& $-$ & $0.4736(7)$ & 
$ 0.4734(7)$\\
 $\delta^{QCD}_{j}$ [\%] & $-$ & $-7.18\pm  0.13$ &  $-7.21\pm 
 0.13$   \\ 
\hline
\end{tabular}
\vspace*{1em}
\label{tab:nlo}
\end{center}
\end{table}

\section{Six fermion production~\cite{lusifer}}

Since top quarks decay via the cascade $t\to b W^+\to b f\bar f'$ into three
fermions, the production of $t\bar t$ pairs corresponds to a particular class
of $e^+e^-\to 6f$ processes: $e^+e^-\to b\bar b  f_1 \bar f'_1 f_2 \bar f'_2$,
where $f_i \bar f'_i$ denote two weak isospin doublets as
shown in Fig.~\ref{fig:ttgraphs}. Ref.~\cite{lusifer}
presents the Monte Carlo event generator {\tt Lusifer}, which 
is designed for all SM processes $e^+e^-\to 6$~fermions in
lowest order.\footnote{Note that Ref.~\cite{Gleisberg:2003bi} describes similar
results based on the HELAC/PHEGAS~\cite{Kanaki:2000ms} and 
AMEGIC++~\cite{Schalicke:2002ck} packages}   
Gluon-exchange diagrams can be included for final
states with two leptons and four quarks (not yet for six-quark final states). 
The matrix elements are evaluated using the Weyl--van~der~Waerden (WvdW) spinor
technique and the phase-space integration is performed using multi-channel
Monte Carlo integration improved by adaptive weight optimization. The
lowest-order predictions are dressed by initial-state radiation (ISR) in the
leading logarithmic approximation following the structure-function
approach~\cite{sf}.

There is a technical problem due to the finite decay widths of unstable
particles in the amplitudes which generates gauge-invariance-breaking effects. 
Already for CM energies in the TeV range these effects are
clearly visible  in some cases, underlining the importance of this issue. 
Within {\tt Lusifer}  several width schemes are implemented, including  
the {\it complex-mass scheme}, which was introduced in Ref.~\cite{Denner:1999gp} for
tree-level predictions and maintains gauge invariance. Hence, gauge-violating
artefacts can be controlled by comparing a given  width scheme with the
complex-mass scheme.

\begin{figure}[tb!]
\centerline{
\setlength{\unitlength}{1pt}
\begin{picture}(155,125)(0,5)
\ArrowLine( 10, 45)( 30, 65)
\ArrowLine( 30, 65)( 10, 85)
\ArrowLine(145, 30)(130, 45)
\ArrowLine(130, 45)(145, 55)
\ArrowLine(145, 75)(130, 85)
\ArrowLine(130, 85)(145,100)
\ArrowLine( 70, 65)(100, 95)
\ArrowLine(100, 35)( 70, 65)
\ArrowLine(100, 95)(145,120)
\ArrowLine(145, 10)(100, 35)
\Photon(30, 65)( 70, 65){2}{6}
\Photon(100, 35)(130, 45){2}{5}
\Photon(100, 95)(130, 85){2}{5}
\Vertex(30,  65){2.0}
\Vertex(130, 45){2.0}
\Vertex(130, 85){2.0}
\Vertex(70, 65){2.0}
\Vertex(100, 35){2.0}
\Vertex(100, 95){2.0}
\put(40,50){$\gamma/ Z$}
\put(80,90){$t$}
\put(80,34){$t$}
\put(105,73){$W$}
\put(105,48){$W$}
\put(150,120){$b$}
\put(150, 5){$b$}
\put( -5, 85){$e^+$}
\put( -5, 40){$e^-$}
\end{picture}
} 
\caption{Diagram for $t\bar t$ production:
$e^+e^-\to t\bar t\to b W^+\bar b W^-\to6f$}
\label{fig:ttgraphs}
\end{figure}

\begin{figure}[tb!]
\begin{center}
\vspace*{0.5cm}
 \epsfig{file=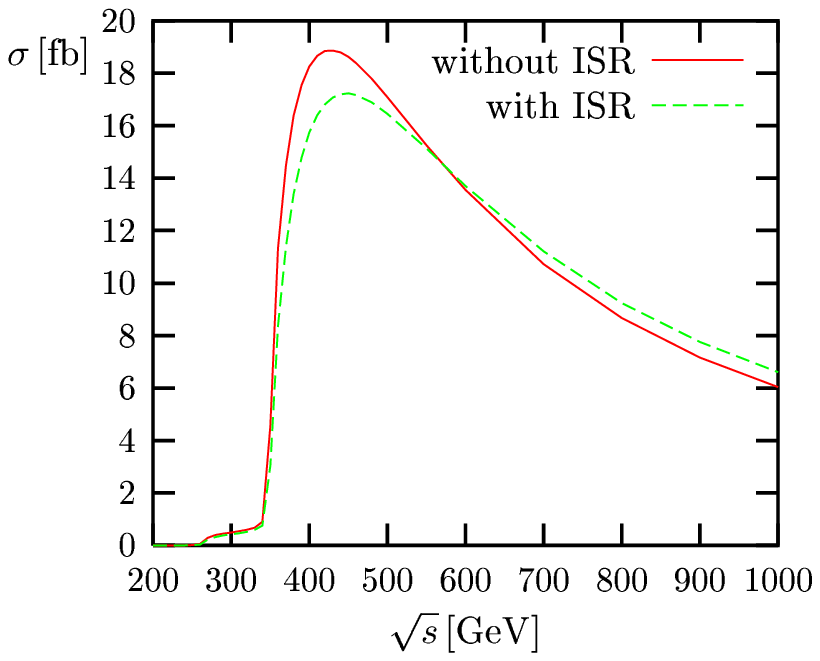,width=5.0cm}
 \epsfig{file=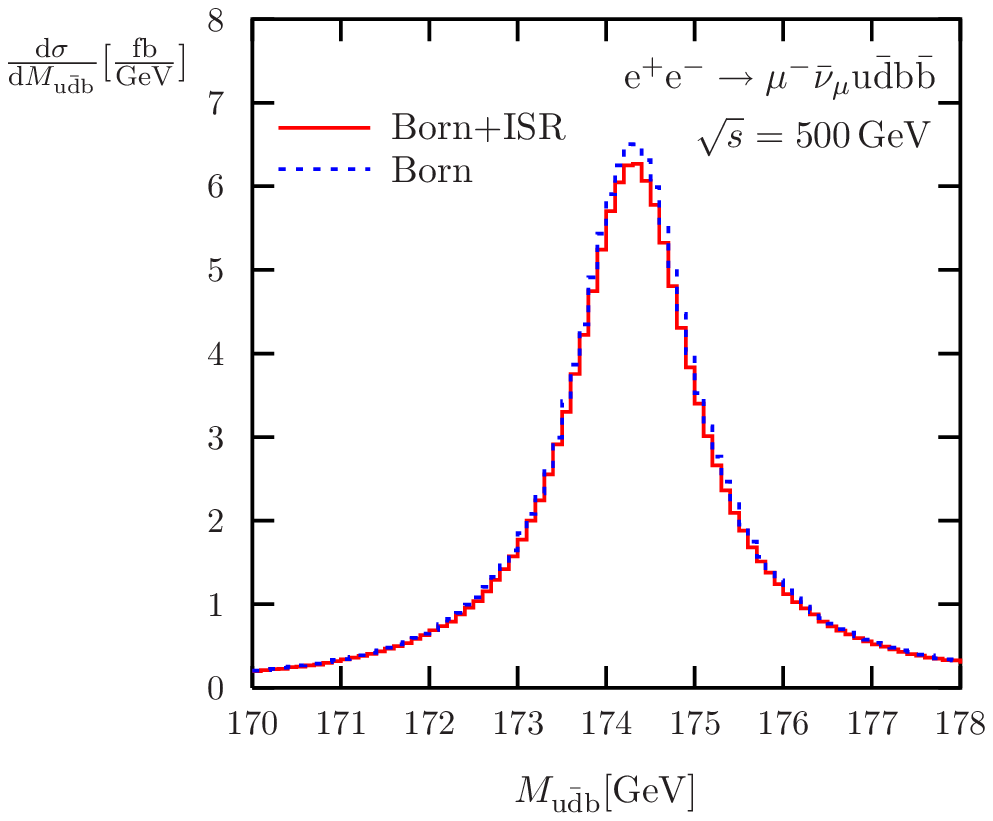,width=5.0cm}
\end{center}
\caption{(a) Total cross section of
$e^+e^-\to\mu^-\bar\nu_\mu u\bar d b\bar b$ (without
gluon-exchange diagrams)
as function of the CM energy with and without ISR and (b) 
Invariant-mass distribution of the $u\bar d b$ quark triplet
in $e^+e^-\to\mu^-\bar\nu_\mu u\bar d b\bar b$ 
(without gluon-exchange diagrams):
absolute prediction with and without ISR}
\label{fig:topmassdist}
\end{figure}

Figure~\ref{fig:topmassdist}(a) illustrates the energy dependence 
of the top-quark pair production cross section for final states where
one of the produced W~bosons decays hadronically and the other
leptonically. 
The cross section steeply rises at the $t\bar t$ threshold, reaches
its maximum between $400$~GeV and $500$~GeV, and then decreases with
increasing energy. We see that ISR reduces the
cross section for energies below its maximum and enhances it above, 
thereby shifting the maximum to a higher energy. This behaviour is
simply due to the radiative energy loss induced by ISR. Near a CM energy
of $250$~GeV the onset of $WWZ$ production can be observed.
Note that this contribution is entirely furnished by background
diagrams, i.e.\ by diagrams that do not have a resonant top-quark
pair.
Figure~\ref{fig:topmassdist}(b) shows the invariant-mass distribution of the 
$u\bar d b$ quark triplet that results from the top-quark decay.
As expected, ISR does not distort the resonance shape but merely
rescales the Breit--Wigner-like distribution.

{\tiny
\begin{table}
\begin{center}
\begin{tabular}{|c|c||c|c|c|}
\hline
\multicolumn{5}{|c|}{
$\sigma(e^+e^-\to\mu^-\bar\nu_\mu u\bar d b\bar b)\,[{\rm fb}]$}
\\ \hline \hline
\multicolumn{2}{|c||}{$\sqrt{s}[{\rm GeV}]$} & 500 & 800 & 2000
\\ \hline \hline
{\tt Lusifer} 
& fixed width /& 17.095(11) & 8.6795(83) & 1.8631(31)
\\[-.2em]
& step width  & & &   
\\ \cline{2-5} 
& running width & 17.106(10) & 8.6988(85) & 2.3858(31)
\\ \cline{2-5} 
& complex mass & 17.085(10) & 8.6773(84) & 1.8627(31)
\\ \hline  \hline 
{\tt W.\&{}M.} & step width & 17.1025(80) & 8.6823(44) & 1.8657(12)
\\ \hline 
\end{tabular}
\caption{Born cross sections (without ISR and gluon-exchange diagrams) for 
$e^+e^-\to\mu^-\bar\nu_\mu u\bar d b\bar b$
for various CM energies and schemes for introducing decay widths}
\label{tab:wwww_width}
\end{center}
\end{table}
}

Table~\ref{tab:wwww_width} shows the effect of using different schemes for
introducing finite decay widths. In spite of  violating gauge invariance, the
fixed width practically yields the same results as the complex-mass scheme that
maintains gauge invariance. Table~\ref{tab:wwww_width} also shows some
results obtained
from the multi-purpose packages {\tt Whizard}~\cite{Kilian:2001qz} and 
{\tt Madgraph}~\cite{Stelzer:1994ta}. In general, and apart from a
few cases, where the limitations of {\tt Whizard} and {\t Madgraph} become
visible, there is good numerical agreement, demonstrating the reliability of
{\tt Lusifer}. 

\section{Top production in the asymptotic regime~\cite{Beccaria:2000jz,
Beccaria:2001an,Beccaria:2002tz,Beccaria:2004yt,Beccaria:2004qg}}

At energies far above the electroweak scale, $\sqrt{s} >\!\!> M \sim M_W\sim
M_Z$, the electroweak corrections are enhanced by large logarithmic corrections
of the type
$$
\alpha^L \log^N\left(\frac{s}{M^2}\right), \qquad 1 \leq N \leq 2L.$$
The leading logarithmic corrections correspond to $N=2L$. These corrections are
related to the singular part of the radiative corrections in the massless limit
$M^2/s \to 0$.  They are either remnants of UV singularities or mass
singularities from soft/collinear emission of virtual or real particles from
initial or final state particles.   This is because the mass of the gauge
bosons provide a physical cut-off to the real radiation.  Furthermore, the
Bloch-Nordsieck theorem is violated for inclusive quantities if the asympototic
states carry non-abelian charges.

The top quark effective vertex of Eq.~\ref{3forms} receives logarithmic
corrections in the asymptotic limit.  In fact, it is possible to see
immediately from the structure of the one-loop Feynman diagrams
of both the SM and the MSSM that the coefficients
of the new extra Lorentz structure $(p-p')^{\mu}$ vanish at large $q^2$ like
$1/q^2$,  while those of the conventional Lorentz structures
($\gamma^{\mu},~\gamma^{\mu}\gamma^5$) can produce either quadratic or linear
logarithms. Therefore, the leading terms of $t\bar t$ production at
asymptotic energies are exactly those that would be computed in a
conventional scheme in which the new scalar component of eq.(\ref{3forms})
has been neglected, and \underline{four} independent gauge-invariant
combinations survive that are, formally, equivalent to those of the final light
quark case.

\begin{figure}[tb!]
\begin{center}
\epsfig{file=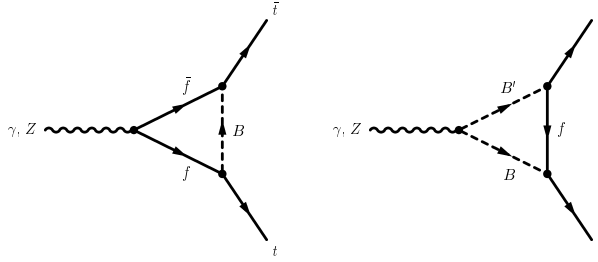,width=8cm}
\end{center}
\caption{Triangle SM diagrams contributing to the asymptotic
logarithmic behaviour in the energy; $f$ represent $t$ or $b$ quarks,
$B$ represent $W^{\pm}$, $\Phi^{\pm}$ or $Z$, $G^{0}$, $H_{SM}$.}
\label{figsm}
\end{figure}

\begin{figure}[tb!]
\begin{center}
\epsfig{file=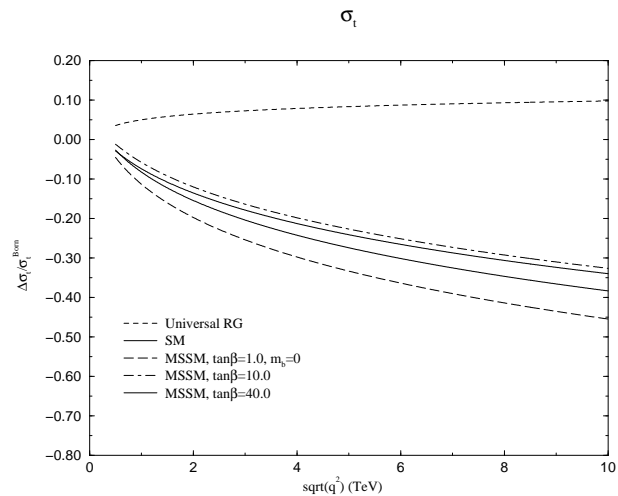,width=8cm}
\end{center}
\caption{Relative effects on the $t\bar t$ cross section $\Delta
\sigma_t/\sigma_t$ in $e^+e^-$
annihilation at CM energy $\sqrt{q^2}$ due to the asymptotic 
logarithmic terms.}
\label{sigt}
\end{figure}

Within the SM, typical diagrams giving rise to these
logarithms are shown in Fig.~\ref{figsm}.
The one-loop logarithmic corrections in the SM
have been computed in Ref.~\cite{Beccaria:2000jz} for
the integrated $e^+e^-\to t\bar t$ cross section, 
$\sigma_t$, 
\begin{eqnarray}
\sigma_{t}&=&\sigma^{B}_{t}\Biggl(1+{\alpha\over4\pi}\Biggl((8.87N-33.16
)\ln{q^2\over\mu^2}+(22.79 \ln{q^2\over M^2_W}-
5.53\ln^2{q^2\over M^2_W})
\nonumber\\
&&
+(3.52\ln{q^2\over M^2_Z}-1.67\ln^2{q^2\over M^2_Z})
-~14.21 \ln{q^2\over m^2_t}
\Biggr)\Biggr),
\label{sigtt}\end{eqnarray}
the forward backward asymmetry $A_{FB,t}$, \begin{eqnarray}
A_{FB,t}&=&A^{B}_{FB,t}+{\alpha\over4\pi}\Biggl((0.45N-4.85
)\ln{q^2\over\mu^2}-(1.79 \ln{q^2\over M^2_W}+0.17\ln^2{q^2\over M^2_W})
\nonumber\\
&&
-(1.26\ln{q^2\over M^2_Z}+0.06\ln^2{q^2\over M^2_Z})
+0.61 \ln{q^2\over m^2_t}
\Biggr)
 \ ,
\label{AFBt}
\end{eqnarray}
the
longitudinal polarization asymmetry $A_{LR,t}$ and its
forward-backward polarization asymmetry $A_t$.
The MSSM effects have also been computed~\cite{Beccaria:2000jz}. 
The effects are largest for the total cross section as shown in Fig.~\ref{sigt}.

The conclusion is that the leading
electroweak effect at the one-loop level is
quite sizeable in the TeV region in all observables, with the only (expected)
exception of the forward-backward asymmetry.
These effects are systematically larger
than those in the corresponding lepton or ``light'' $(u,d,s,c,b)$ quark
production observables, both in the SM and MSSM.
In the latter case, top production exhibits also in the
leading terms a strong dependence on $\tan\beta$, much stronger than
that of bottom production.

In the asymptotic region, the different  effects on the $t\bar t$ and $b\bar b$
cross sections can in principle be
exploited~\cite{Beccaria:2001an,Beccaria:2002tz,Beccaria:2004yt,Beccaria:2004qg}.   
Refs.~\cite{Beccaria:2004qg} examine the effect at the LHC under the assumption
of a ``moderately'' light SUSY scenario and find that   the  electroweak and
the strong SUSY contributions combine to produce an enhanced effect whose
relative value in the $t\bar t$ and $b\bar b$ cross sections could be as large
as 20\%  for large values of $\tan\beta$.


\section{Top quark couplings}

\subsection{$Wtb$~\cite{delAguila:2002nf}}

The most general CP-conserving $Wtb$ vertex can be parameterised
with the effective Lagrangian 
given by \footnote{The most general $Wtb$ vertex (up to dimension five)
involves ten operators, but at the expected level of precision it is an
excellent approximation to consider the top on-shell. With $b$ also on-shell
and $W \to l \nu,jj$ six of them can be eliminated using Gordon identities.
The resulting Lagrangian can be further restricted assuming CP conservation.
The couplings can then be taken to be real, of either sign.}
\begin{eqnarray}
\mathcal{L} & = & - \frac{g}{\sqrt 2} \bar b \, \gamma^{\mu} \left( V_{tb}^{L}
P_L + V_{tb}^R P_R
\right) t\; W_\mu^- \nonumber \\
& & - \frac{g}{\sqrt 2} \bar b \, \frac{i \sigma^{\mu \nu} q_\nu}{M_W}
\left( g^L P_L + g^R P_R \right) t\; W_\mu^- + \mathrm{h.c.} 
\label{ec:1}
\end{eqnarray}
In the SM the $Wtb$ vertex is purely left-handed and its size
is given by the Cabibbo-Kobayashi-Maskawa (CKM) matrix element $V_{tb}^L
\equiv V_{tb}$. The right-handed vector and both tensor couplings vanish at
tree-level in the SM, but can be generated at higher orders in the SM or its
extensions \cite{Beneke:2000hk}. Note that $V_{tb}^R$ is constrained by $b \to
s\gamma$ decays while  the $\sigma^{\mu\nu}$ terms are not because
of the extra $q^{\mu}$ factor that suppresses their effect in $b$
decays. The $Wtb$ vertex structure can be probed and measured using either
top-pair production or single-top-quark production processes.   The $t \bar t$
cross-section is rather insensitive to the size
of $V_{tb}$ and 
to obtain a measure of the {\em absolute} value of $V_{tb}$ it is necessary to
fall back on less abundant single top production \cite{papiro2}, with a rate
proportional to $|V_{tb}|^2$.  Nevertheless, $t\bar t$ production can give
invaluable information on the $Wtb$ vertex. Angular asymmetries between decay
products are very sensitive to a small admixture of a right-handed
$\gamma^{\mu}$ term or a $\sigma^{\mu \nu}$ coupling of either chirality.

In Ref.~\cite{delAguila:2002nf}, the forward-backward asymmetry
 in the decay of the top quark $t \to
W^+ b \to l^+ \nu b$ as measured in the $W$ rest frame
is proposed as a particularly sensitive probe of
anomalous top quark couplings. It is defined as
\begin{equation}
A_\mathrm{FB} = \frac{N(x_{bl} > 0) -
N(x_{bl} < 0)}{N(x_{bl} > 0) +
N(x_{bl} < 0)}\,,
\end{equation}
where $x_{bl}$ is the cosine of the angle between the 3-momenta of the $b$
quark and the charged lepton in the $W$ rest frame, and $N$ stands for the
number of events. The same definition holds for the $\bar t \to l^- \bar \nu
\bar b$ decay.

\begin{figure}[tb!]
\vspace{5mm}
\begin{center}
\mbox{\epsfig{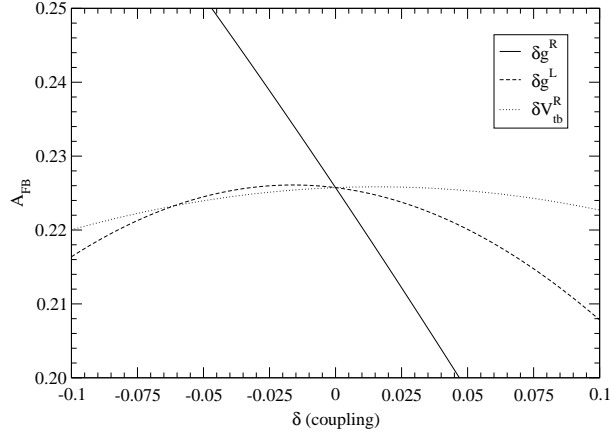}}
\end{center}
\caption{Dependence of $A_\mathrm{FB}$ on $\delta g^R$ (solid line),
$\delta g^L$ (dashed line) and $\delta V_{tb}^R$ (dotted line).
The SM result occurs where all three lines cross. 
We use $m_t = 175$, $M_W =
80.33$, $m_b = 4.8$ GeV.
\label{fig:afb} }
\end{figure}

$A_\mathrm{FB}$ only depends on the $t$, $b$ and $W$ boson masses, and on the
couplings in Eq.~(\ref{ec:1}).  The SM tree-level (LO) value is $A_\mathrm{FB} =
0.2223$ while the bulk effect of the 
one-loop QCD corrections can be
taken into account by including a $\sigma^{\mu \nu}$ term $g^R = -0.00642$
\cite{papiro8}. The corresponding NLO value is $A_\mathrm{FB} = 0.2257$.  In
Fig.~\ref{fig:afb} we plot $A_\mathrm{FB}$ for different values of $\delta g^R
\equiv g^R + 0.00642$, $\delta g^L \equiv g^L$ and $\delta V_{tb}^R \equiv
V_{tb}^R$. 
Numerical studies of the tree-level $2 \to 6$ processess $gg,\;q \bar q \to t
\bar t \to W^+ b W^- \bar b \to l \nu jjjj$ plus $Wjjjj$ background,
including all spin correlations
and realistic cuts suggest that a statistical error of $\delta A_\mathrm{FB}
\simeq 5 \times 10^{-4}$ is achievable at the LHC.   The
main systematic errors come from the uncertainty in $m_t$ and $M_W$ and will
be negligible with ILC precision.

The cross sections in the forward and backward hemispheres are of the order of
11-16~pb (2-3~pb) for the signal (background). Using both electron and muon
channels leads to a (statistical) sensitivity of
$\delta g^R =\pm 0.003$, $\delta g^L = +0.02 (-0.05)$ and 
$\delta V_{tb}^R=+0.08 (-0.04)$.
The sensitivity to $g^R$ is one order of magnitude better than
in single top production at LHC \cite{papiro7} while the
sensitivity to $g^L$ is competitive with that expected at the ILC, 
or from single top production at LHC. 

\subsection{Flavour Changing Neutral
Couplings~\cite{Bejar:2000ub,Bejar:2001sj,Aguilar-Saavedra:2001ab,Aguilar-Saavedra:here}}

Flavor Changing Neutral (FCN) decays of the top quark within the strict
context of the Standard Model are  known to be extremely rare.  In fact, there
are no tree-level FCN current   processes in the
Standard Model. However, they can be generated at the one-loop level by
charged current interactions.  The most general effective Lagrangian
describing the possible interactions of a top quark, a light quark $q$ and a
$Z$ boson, photon $A$, gluon $G$ or Higgs $H$ can be written as,
\begin{eqnarray}
{\mathcal L} & = & -\frac{g^\prime}{2} \, X_{tq} \, \bar t \gamma_\mu  
(x_{tq}^L P_L - x_{tq}^R P_R) q Z^\mu 
- \frac{g^\prime}{2} \, \kappa_{tq}\, \bar t (\kappa_{tq}^{v}+ \kappa_{tq}^{a}
\gamma_5) \frac{i \sigma_{\mu \nu} q^\nu}{m_t} q Z^\mu  \nonumber \\
& &  - e \, \lambda_{tq}\, \bar t (\lambda_{tq}^{v}+ \lambda_{tq}^{a} \gamma_5)
\frac{i \sigma_{\mu \nu} q^\nu}{m_t} q A^\mu 
-g_s\zeta_{tq}\bar t  \left(\zeta_{tq}^V+\zeta_{tq}^A\gamma_5\right) 
\frac{i \sigma_{\mu \nu} q^\nu}{m_t} q T^aG^{a\mu}\nonumber \\
&&-\frac{g}{2\sqrt{2}}g_{tg}\bar t \left(g_{tq}^V+g_{tq}^Q\gamma_5\right)q H + h.c.\,, 
\label{Lfcnc}
\end{eqnarray}
where $g^\prime = g/\cos\theta_W$, $P_{R,L}=(1 \pm \gamma_5)/2$.  The chirality-dependent couplings are
constants and are normalized to $(x_{tq}^L)^2+(x_{tq}^R)^2=1$ etc.  Within the
SM, all of these vertices vanish at the tree-level, but can be generated
at one-loop by charged current interactions.   However, because of GIM
cancellations, the one-loop effects are parametrically suppressed beyond naive
expectations based on pure dimensional analysis,  power counting and CKM
matrix elements by $m_b^4/M_W^4$. The Standard Model  typical branching ratios
for these rare top quark decays   are so small ($\sim 10^{-12}$--$10^{-17}$)
that they are hopelessly undetectable at the Tevatron, LHC and ILC in any of
their scheduled upgradings. 

\begin{figure}[tb!]
\begin{center}
\epsfig{figure=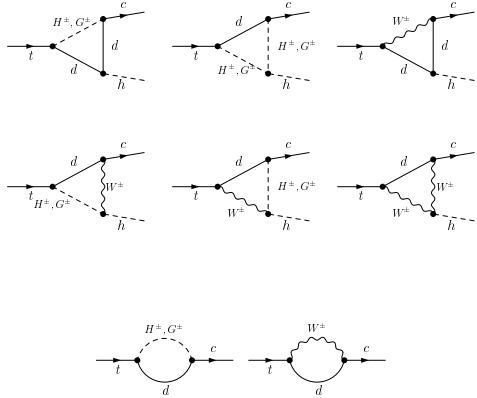, height=6cm}
\caption{One-loop vertex diagrams contributing to the FCN top quark decays. 
Shown are the vertices and mixed self-energies with 
all possible contributions from the SM fields and the Higgs bosons from the 
general 2HDM. } 
\label{fig:FCNC}
\end{center}
\end{figure}

Refs.~\cite{Bejar:2000ub,Bejar:2001sj} considered loop induced $ t\to ch$ and
$t \to cg$ FCN decays in the MSSM and in the general two-Higgs doublet model
(2HDM) - see also Ref.~\cite{Guasch:1999jp}.   
Typical diagrams contributing to these decays are shown in
Fig.~\ref{fig:FCNC}. The 2HDM parameter space is constrained by the $\rho$
parameter and the one-loop corrections to the $\rho$-parameter from the 2HDM
sector cannot  deviate from the reference SM contribution by more than one per
mille,   $|\delta\rho^{2HDM}|< 0.001$. There are also constraints
on the charged Higgs from radiative $B$ decays.  Nevertheless the  fiducial
branching ratio defined by 
\begin{equation} 
B^{j}(t\rightarrow X+c)=\frac{\Gamma^{j}(t\rightarrow X+c)}{\Gamma 
(t\rightarrow W^{+}+b)+\Gamma^{j}(t\rightarrow H^{+}+b)}\,\,, 
\label{fiducialH} 
\end{equation} 
may be as large as $10^{-5}$ for top decay into the lightest CP-even higgs and
$10^{-6}$ for $t \to g c$. Values for other models are reviewed in
~\cite{Aguilar-Saavedra:here}.\footnote{
Note that the related decay $H \to t\bar c$
is discussed in the context of the 2HDM in~\cite{Bejar:2003em}.  
The isolated top quark signature, unbalanced by any other heavy particle 
should help to identify the FCN event and makes branching ratios of 
$10^{-5}$ accessible at the LHC.}

\begin{figure}[tb!]
\begin{center}
\epsfig{figure=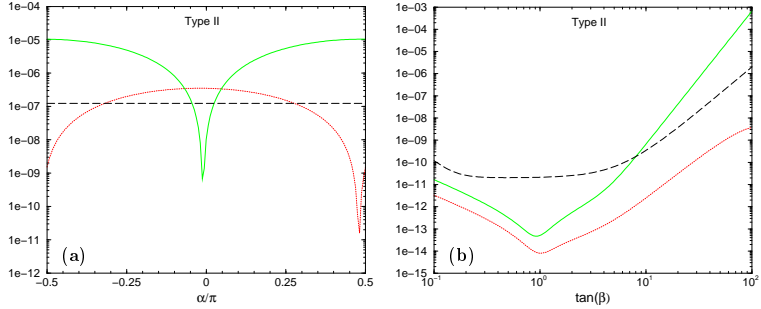, width=10cm}
\caption{Evolution of the FCNC top quark fiducial ratios in Type~II 2HDM 
as functions of (a) the mixing angle $\alpha$ in the CP-even Higgs sector, 
and (b)  $\tan\beta$ for $t \to Xc$ with $X=h$ (green), $X-H$ (red) and $X=g$ (dash).
The plot in (b) 
continues above the usual bound on $\tan\beta$ just to better show the 
general trend.} 
\label{fig:2hdm2}
\end{center}
\end{figure}

To illustrate the potential effects in a 2HDM model,
Fig.~\ref{fig:2hdm2} shows  the branching ratios for $t\to Xc$ for
$X=h,H,~g$ as functions of the parameters $\alpha$ and $\beta$. The 
highest potential rates are of order $10^{-5}$, and so there is hope
for  being visible.

Current limits on FCN top decays from the Tevatron, LEP and HERA are at the
few per cent level.  Run 2 at the Tevatron is expected to reduce these limits
by about an order of magnitude. At the LHC, the search for FCN top couplings
can be carried out examining two different types of processes. On the one
hand, we can look for FCN top decays in  $gg,q \bar q \to t \bar t \to X q W
b$ where $X=\gamma, Z, g$ or Higgs.   On the other hand, one can search for
single top production via an anomalous effective vertex such as $qg \to X t$  
where the top quark is assumed to decay in the SM dominant mode $t \to Wb$. 
The main backgrounds are thus $t\bar t$, $W + {\rm jets}$, $VV+{\rm jets}$ and
single top production.    Numerical simulations of signal and background
indicate that the LHC will improve by at least a factor of 10 on the Tevatron
sensitivity to around $10^{-5}$.

\begin{figure}[b!]
\begin{center}
\mbox{\epsfig{file=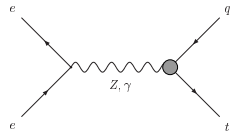,width=3cm}}
\end{center}
\caption{Feynman diagrams for $e^+ e^- \to t\bar q$ via $Ztq$ or $\gamma tq$
FCN couplings. The top quark is off-shell and decays to $Wb$.
\label{fig:feyn1} }
\end{figure}

At the ILC, the top pair production cross section is much smaller  than at the
LHC and  the limits obtained from top decays cannot compete with those from
the LHC. The capabilities of the ILC have been studied in
Ref.~\cite{Aguilar-Saavedra:2001ab} for the single top production processes
$e^+e^- \to tq$ shown in Fig.~\ref{fig:feyn1} and $e^+e^- \to tq\gamma$ and
$e^+e^- \to tqZ$.  The  signal matrix elements including the top decay were
evaluated using HELAS \cite{papiro18} and introducing a new HELAS-like
subroutine {\tt IOV2XX} to compute the non-renormalizable $\sigma_{\mu \nu}$
vertex. The relevant backgrounds are $e^+ e^- \to W^+ q \bar q'$, $W^+ q \bar
q'Z$ and $W^+ q \bar q'\gamma$ and were evaluated using MadGraph
\cite{Stelzer:1994ta}.  

Assuming one year of running time in all the cases, that
is, 100 fb$^{-1}$ for LHC, 300 fb$^{-1}$ for ILC at 500 GeV 
and no beam polarisation,
 Refs.~\cite{Aguilar-Saavedra:2000db,Aguilar-Saavedra:2001ab} find that by combining
the information from both production and decay, 
the sensitivities on the $t \to Xc$ coupling 
are given in Table~\ref{tab:lim}.
The most optimistic case with 500 fb$^{-1}$ of data 
80\% polarised electron and 60\% polarised positron
beams and a CM energy of 800~GeV is denoted by ILC+.
\begin{table}[htb]
\begin{center}
\begin{tabular}{|c|c|c|c|}
\hline
&  {LHC} &  {ILC}&  {ILC+} \\
\hline
$\mathrm{Br}(t \to Zc)$ $(\gamma_\mu)$ &
   $3.6 \times 10^{-5}$ &
   $1.9 \times 10^{-4}$ &
   $1.9 \times 10^{-4}$ \\
$\mathrm{Br}(t \to Zc)$ $(\sigma_{\mu \nu})$ &
   $3.6 \times 10^{-5}$ &
   $1.8 \times 10^{-5}$ &
   $7.2 \times 10^{-6}$ \\
$\mathrm{Br}(t \to \gamma c)$ &
   $1.2 \times 10^{-5}$ &
   $1.0 \times 10^{-5}$ &
   $3.8 \times 10^{-6}$ \\ \hline
\end{tabular}
\end{center}
\caption{3$\sigma$ discovery limits on top FCN couplings that can be obtained at LHC and ILC
for one year of operation.
\label{tab:lim}}
\end{table}

We see that LHC and ILC complement each other in the search for top FCN
vertices. The $\gamma_\mu$ couplings to the $Z$ boson can be best measured or
bound at LHC, whereas the sensitivity to the $\sigma_{\mu \nu}$ ones is better
at ILC. 
For a more detailed discussion, see Ref.~\cite{Aguilar-Saavedra:here}

\section{Impact of a precise top mass measurement~\cite{Heinemeyer:2003ud,Heinemeyer:2004ju}}

The current world average for the top-quark mass is $m_t = 178.0 \pm
4.3$~GeV~\cite{Azzi:2004rc,natured0}.  The expected accuracy at the 
Tevatron and the LHC is $\delta m_t = \mbox{1--2}$~GeV~\cite{Beneke:2000hk},  while
at the ILC a very precise determination of $m_t$ with an accuracy of $\delta m_t
\lsim 100 ~{\rm MeV}$ should be 
possible~\cite{Aguilar-Saavedra:2001rg,Abe:2001wn,Abe:2001gc,mtdet}. This error contains both
the experimental error of the mass parameter extracted from the $t \bar t$
threshold measurements at the ILC and  the expected theoretical uncertainty
from its transition into a suitable short-distance mass (like the \msbar\
mass).

\subsection{Electroweak Precision Observables}

Electroweak precision observables (EWPO) can be used to perform internal
consistency checks of the model under consideration and to obtain indirect
constraints on unknown model parameters. This is done by comparing experimental
results for the precision observables with their theory prediction within, for
example, the Standard Model (SM). Any improvement in the precision of the
measurement of $m_t$ will have an effect on the analysis of EWPO of which the
two most prominent are the $W$~boson mass $M_W$ and the effective leptonic
mixing angle $\sin^2\theta_{{\rm eff}}$. 

Currently the uncertainty in $m_t$ is by far the dominant effect in the
theoretical uncertainties of the EWPO. Today's experimental errors of $M_W$ and
$\sin^2\theta_{{\rm eff}}$~\cite{ewdataw03} are shown in Table~\ref{tab:ewpounc},
together with the prospective future experimental errors at high energy
colliders (see \cite{blueband} for a compilation of
these errors and additional references).

\begin{table}[htb!]
\renewcommand{\arraystretch}{1.5}
\begin{center}
\begin{tabular}{|c||c|c|c|c|}
\cline{2-5} \multicolumn{1}{c||}{}
& Today & Tevatron/LHC & ~ILC~  & GigaZ \\
\hline\hline
$\delta \sin^2\theta_{{\rm eff}}(\times 10^5)$ & 16 & 14--20   & --  & 1.3  \\
\hline
$\delta M_W$ [MeV]           & 34 & 15   & 10   & 7      \\
\hline\hline
\end{tabular}
\end{center}
\caption{Experimental errors of $M_W$ and $\sin^2\theta_{{\rm eff}}$ at present and future
  colliders~\cite{ewdataw03,blueband}. 
}
\label{tab:ewpounc}
\renewcommand{\arraystretch}{1}
\end{table}

In general, there are two sources of theoretical uncertainties: those from
unknown higher-order corrections (``intrinsic'' theoretical uncertainties), and
those from experimental errors of the input parameters (``parametric''
theoretical uncertainties). The intrinsic uncertainties within the SM are 
\BE
\Delta M_W^{\rm intr,today} \approx 4~{\rm MeV},\qquad\qquad 
\Delta \sin^2\theta_{{\rm eff}}^{\rm intr,today} \approx 4.9 \times
10^{-5}
\label{eq:intruncSM}
\EE 
at present~\cite{mwsweff,mwsweff2}.  
They are based on the present status of the theoretical predictions in
the SM, namely the complete two-loop result for $M_W$ (see
\cite{mwsweff,MWtwoloop} and references therein),
the complete two-loop fermionic result for $\sin^2\theta_{{\rm eff}}$ 
(see \cite{mwsweff2}, previous partial results and references can be found in~\cite{dgs})
 and leading three-loop contributions to both observables (see
\cite{faisst} for the latest result, and references therein). 

The current parametric uncertainties induced by the experimental errors of
$m_t$~\cite{MWradcor} are
\BE
\delta m_t = 4.3 ~{\rm GeV} \Rightarrow 
\Delta M_W^{{\rm para},m_t} \approx \pm 26 ~{\rm MeV}, \quad
\Delta \sin^2\theta_{{\rm eff}}^{{\rm para}, m_t} \approx \pm 14 
\times 10^{-5}.\EE
We see that the parametric uncertainties of $M_W$ and $\sin^2\theta_{{\rm eff}}$
induced by $\delta m_t$ are
approximately as large as the current experimental errors.\footnote{
Note that the parametric errors induced by $\delta(\Delta\alpha_{\rm had})$
are 
$\Delta M_W^{{\rm para},\Delta\alpha_{\rm had}} \approx \pm 6.5 ~{\rm MeV}$
and  $\Delta \sin^2\theta_{{\rm eff}}^{{\rm para},\Delta\alpha_{\rm had}} \approx 
  \pm 13 \times 10^{-5}$~\cite{MWradcor}}

A future experimental error of $\delta m_t \approx
1.5$~GeV at the LHC will give rise to parametric uncertainties of $$\Delta
M_W^{\rm para,LHC} \approx 9 ~{\rm MeV},\qquad\qquad\Delta \sin^2\theta_{{\rm eff}}^{\rm para,LHC}
\approx 4.5 \times 10^{-5}.$$ 
On the other hand,  the  ILC precision of $\delta m_t \approx
0.1$~GeV will reduce the parametric uncertainties to $$\Delta M_W^{\rm para,ILC}
\approx 1 ~{\rm MeV},\qquad\qquad\Delta \sin^2\theta_{{\rm eff}}^{\rm para,ILC} \approx 0.3 \times
10^{-5}.$$ 
In order to keep the theoretical uncertainty induced by $m_t$ at 
a level comparable to or smaller than the other parametric and intrinsic
uncertainties, $\delta m_t$ has to be smaller than about $0.2 ~{\rm GeV}$ in the
case of $M_W$, and about $0.5 ~{\rm GeV}$ in the case of $\sin^2\theta_{{\rm eff}}$.
In other words, ILC accuracy on $m_t$ will be
necessary in order to keep the parametric error induced by $m_t$ at or below
the level of the other uncertainties. With the LHC accuracy on $m_t$, on the
other hand, $\delta m_t$ will be the dominant source of uncertainty.

\begin{figure}[tb!]
\begin{center}
\epsfig{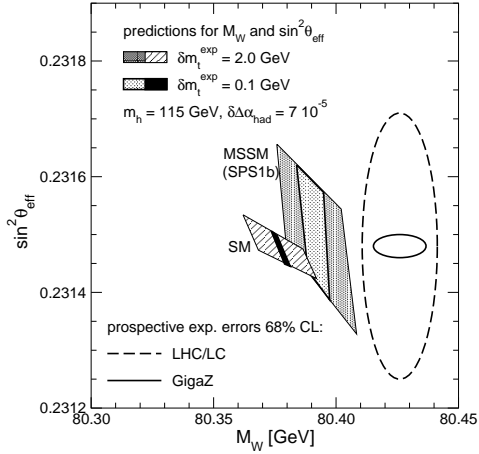}
\caption{
The predictions for $M_W$ and $\sin^2\theta_{{\rm eff}}$ in the SM and MSSM. The inner 
(blue) area corresponds to $\delta m_t^{exp} = 0.1 ~{\rm GeV}$ (ILC), while
the outer (green) area arises from $\delta m_t^{exp} = 2 ~{\rm GeV}$ (LHC). 
The anticipated experimental errors on $M_W$ and $\sin^2\theta_{{\rm eff}}$ at the
LHC/ILC and at an ILC with GigaZ option are indicated. 
}
\label{fig:SWMW}
\end{center}
\end{figure}

As an example of the potential of a precise measurement of the EWPO
to explore the effects of new physics, Fig~\ref{fig:SWMW} shows the
predictions for $M_W$ and $\sin^2\theta_{{\rm eff}}$ in the SM in
comparison with the prospective experimental accuracy obtainable at
the LHC and the ILC without GigaZ option (labelled as LHC/ILC) and with the 
accuracy obtainable at an ILC with GigaZ option (labelled as GigaZ).
The current experimental values are taken as the central ones~\cite{ewdataw03}.
For the Higgs boson mass a future measured value of $m_h = 115~{\rm GeV}$
has been assumed (in accordance with the final lower bound obtained at
LEP~\cite{mhLEPfinal}).
We see that the improvement in $\delta m_t$ from
$\delta m_t = 2 ~{\rm GeV}$ to $\delta m_t = 0.1 ~{\rm GeV}$ 
strongly reduces the
parametric uncertainty in the prediction for the EWPO and 
leads to a reduction by about a factor of 10
in the allowed parameter space of the $M_W$--$\sin^2\theta_{{\rm eff}}$ plane. 


\subsection{Indirect determination of the SM top Yukawa coupling}
\label{subsec:topyuk}

A high precision on $m_t$ is also important to obtain indirect
constraints on the top Yukawa coupling $y_t$ from EWPO~\cite{ytdet}.
The top Yukawa coupling enters the SM 
prediction of EWPO starting at ${\cal O}(\alpha\alpha_t)$~\cite{delrhoSMal2}.
Indirect bounds on this coupling can be obtained if one assumes that
the usual relation between the Yukawa coupling and the top quark mass,
$y_t = \sqrt{2} m_t/v$ (where $v$ is the vacuum expectation value), is
modified. 

Assuming a precision of $\delta m_t = 2 ~{\rm GeV}$, an indirect determination
of $y_t$ with an accuracy of only about 80\% can be obtained from the EWPO
measured at an LC with GigaZ option. A precision of $\delta m_t = 0.1 ~{\rm GeV}$, 
on the other hand, leads to an accuracy of the indirect determination of 
$y_t$ of about 40\% which is competitive with the
indirect constraints from the $t \bar t$~threshold~\cite{mtdet2}. 
These indirect determinations of $y_t$ represent an independent 
and complementary approach to the direct measurement of $y_t$ via 
$t \bar t H$ production at the ILC, which of course provides the
highest accuracy~\cite{Aguilar-Saavedra:2001rg}. 

\subsection{The MSSM}

Within the MSSM, EWPO are also heavily
influenced by the accuracy of the top quark mass. However, the available results beyond one-loop order are less advanced than
in the SM (for the latest two-loop results, see \cite{ytdet} 
and references therein). Thus, the intrinsic uncertainties in the MSSM
are still considerably larger than the ones quoted for the SM in 
Eq.~\ref{eq:intruncSM}.
Fig.~\ref{fig:SWMW} also shows the predictions for $M_W$ and $\sin^2\theta_{{\rm eff}}$
in the MSSM where the MSSM parameters have
been chosen in this example according to the
reference point SPS~1b~\cite{sps}, and all SUSY parameters have been
varied within realistic error intervals. 
In the MSSM case, where many additional parametric uncertainties enter,
a reducing $\delta m_t$ from $2$~GeV to 0.1~GeV leads to
reduction in the allowed parameter space of the $M_W$--$\sin^2\theta_{{\rm eff}}$ plane
by a factor of more than 2.

Because of the additional symmetry of the MSSM, a precise knowledge
of $m_t$ yields additional constraints.
For example,  and in contrast to the SM, where the Higgs boson mass 
is a free input parameter,
the mass of the lightest $CP$-even Higgs boson in the MSSM can be 
predicted in terms of other parameters of the model. Thus, precision
measurements in the Higgs sector of the MSSM have the potential to play a 
similar role as the ``conventional'' EWPO for constraining the parameter
space of the model and possible effects of new physics.

\begin{figure}[tb!]
     \begin{center}
     \begin{tabular}{cc}
     \mbox{\epsfig{file=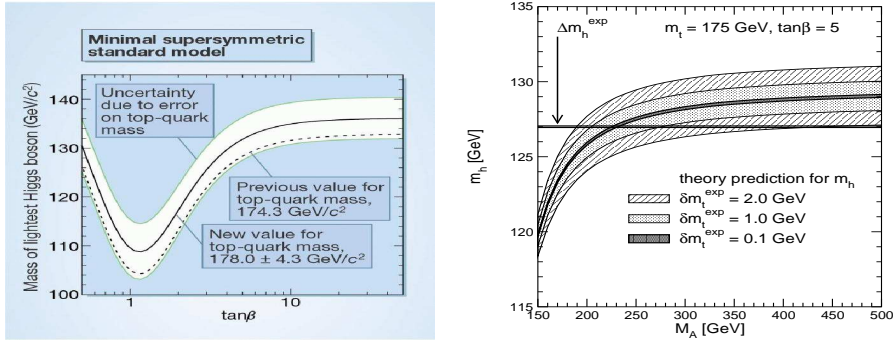,width=5.7cm,height=4.5cm}}&
     \mbox{\epsfig{file=mhMA06.bw.eps,width=5.7cm,height=4.5cm}}
     \end{tabular}
     \end{center}
\vspace{-0.5em}
\caption{Prediction for $m_h$ in the $m_h^{max}$ scenario of the MSSM as a
function of $\tan\beta$ (left) and
$M_A$ (right). In the left plot ~\cite{naturetop}
the impact of the present experimental error of $m_t$ on the $m_h$
prediction is shown. The three bands in the right plot~\cite{Heinemeyer:2003ud}
correspond to $\delta m_t = 1, 2$~GeV (LHC) and $\delta m_t = 0.1$~GeV (ILC).
The anticipated experimental error on $m_h$ at the ILC
is also indicated.
}
     \label{fig:mhmssm}
\vspace{-0.5em}
     \end{figure}

Fig.~\ref{fig:mhmssm} shows the impact of the experimental error of $m_t$ on
the prediction for $m_h$ in the MSSM. The parameters are chosen
according to the $m_h^{max}$ benchmark scenario~\cite{benchmark}. The band
in the left plot corresponds to the present
experimental error of $m_t$~\cite{Azzi:2004rc,natured0}, while in the 
right plot the situation at the LHC ($\delta m_t = 1, 2$~GeV) is compared 
to the ILC
($\delta m_t = 0.1 $~GeV). The figure shows that the ILC precision on
$m_t$ will be necessary in order to match the experimental precision of
the $m_h$ determination with the accuracy of the theory prediction
(assuming that the intrinsic theoretical uncertainty can be reduced to
the same level, see Ref.~\cite{mhiggsAEC}).

Further examples of the importance of 
a precise determination of $m_t$ in the MSSM are
the prediction of sparticle masses, parameter
determinations, and the reconstruction of the supersymmetric high scale
theory~\cite{Heinemeyer:2003ud}.



\section{Other topics}

Other topics of relevance to top quark physics are discussed in the Higgs and 
Electroweak reviews~\cite{Higgs,EW}.

The SM Higgs boson can be searched for in the channels $p \bar p / pp \to
t \bar t H
+ X$ at the Tevatron and the LHC. The cross sections for these processes and the
final-state distributions of the Higgs boson and top quarks are presented at
next-to-leading order QCD in Refs.~\cite{Beenakker:2002nc,Beenakker:2001rj}. 
The impact of the
corrections on the total cross sections is characterized by $K$ factors, the
ratio of next-to-leading order and leading order cross sections.
At the central scale $\mu_0 = (2 m_t + M_H)/2$, the K factors are found to be
slightly below unity for the Tevatron ($K \sim 0.8$) and slightly above unity for
the LHC ($K \sim 1.2$). Including the corrections significantly stabilizes the
theoretical predictions for total cross sections and for the distributions in
rapidity and transverse momentum of the Higgs boson and top quarks.

The two-loop corrections to the heavy quark form factor are studied in
Ref.~\cite{Bernreuther:2004ih} where
 closed analytic expressions of the electromagnetic vertex form
factors for heavy quarks at the two-loop level in QCD are presented
for arbitrary momentum transfer.   This calculation represents
a first step towards 
the two-loop QCD corrections to $t\bar t$ production in both electron-positron annihilation
and hadron collisions.

\section{Summary and Outlook}

There has been significant progress in the study of top quark physics at
current and future particle colliders during the past four years.   As detailed above,
the network 
has contributed to an improved knowledge of the top quark
production and decay properties, both within and without the SM.   
However, much work 
remains to be carried out.
In particular, although the one-loop strong and weak corrections to the top-pair
production cross section are well known, the two-loop QCD corrections are needed to
match the experimental accuracy.   Similarly, it may be necessary to make more
precise predictions of the single top cross section.   Experimental studies of the 
observability of FCN decays of the $t \to Hc$ and $t\to gc$ 
decays are also needed.   

Finally, we  note that the treatment of unstable particles close to resonance
suffers  from the breakdown of ordinary perturbation theory.   A toy model
showing how to systematically improve the calculational accuracy order by order
in perturbation theory  has recently been
proposed~\cite{Beneke:2003xh,Beneke:2004km}.   We anticipate that application
of this improved theoretical approach to the $t\bar t$ cross section close top
threshold should yield an even more accurate experimental determination of the
top quark mass at the ILC.   Because of the large sensitivity of the Higgs
boson mass to $m_t$, this will have an inevitable knock on in any model where
the Higgs mass can be predicted from the other parameters of the theory.

\end{document}